\documentclass[12pt,preprint]{aastex}
\usepackage{graphicx}

\shorttitle{Flux and Spectral Index Correlation}
\begin{document}


\title{Modeling Emission from the Supermassive Black Hole in the Galactic Center with GRMHD Simulations}


\author{Kyle W. Martin,\altaffilmark{1} Siming Liu,\altaffilmark{2} Chris Fragile,\altaffilmark{3}  Cong Yu,\altaffilmark{4} and Chris L. Fryer\ \altaffilmark{2, 5}
}

\altaffiltext{1}{Physics Department, University of New Mexico, Albuquerque, NM 87131; kmartin1@unm.edu}
\altaffiltext{2}{Los Alamos National Laboratory, Los Alamos, NM 87545; 
liusm@lanl.gov}
%\altaffiltext{3}{Physics Department and Steward Observatory, The 
%University of Arizona, 
%Tucson, AZ 85721; melia@physics.arizona.edu; Sir Thomas Lyle Fellow and 
%Miegunyah Fellow.}
\altaffiltext{3}{Physics \& Astronomy, College of Charleston, Charleston, SC 29424; fragilep@cofc.edu}
\altaffiltext{4}{National Astronomical Observatories/Yunnan Observatory, Chinese Academy of Science, Kunming, Yunnan 650011, China; yccit@yahoo.com.cn}
\altaffiltext{5}{Physics Department, The University of Arizona, Tucson, AZ 
85721; fryer@lanl.gov}


\begin{abstract}

Sagittarius (Sgr) A* is a compact radio source at the Galactic center, powered by accretion of fully ionized plasmas into a supermassive black hole of $\sim (3-4) \times 10^6 M_{\bigodot}$. However, the radio emission cannot be produced through the thermal synchrotron process by a gravitationally bounded flow \citep{Liu01}. General relativistic magneto-hydrodynamical (GRMHD) simulations of black hole accretion show that there are strong unbounded outflows along the accretion. With the flow structure around the black hole given by GRMHD simulations, we investigate whether thermal synchrotron emission from these outflows may account for the observed radio emission and discuss the implications of this study on these GRMHD simulations and possible production of non-thermal particles by this source. We find that simulations producing relatively high values of plasma $\beta$ cannot produce the radio flux level without exceeding the X-ray upper limit set by {\it Chandra} observations through the bremsstrahlung process. The predicted radio spectrum is also harder than the observed spectrum both for the one temperature thermal model and a simple nonthermal model with a single power-law electron distribution. Since higher frequency emission is produced at smaller radii, the electron temperature needs to be lower than the gas temperature near the black hole to reproduce the observed radio spectrum. A more complete modeling of the radiation processes, including the general relativistic effects and transfer of polarized radiation, will give more quantitative constraints on physical processes in Sgr A* with the current multi-wavelength, multi-epoch, and polarimetric observations of this source.

\end{abstract}


%% subject header.  Objects should be in the appropriate "individual"
%% headers (e.g. quasars: individual, stars: individual, etc.) with the
%% additional provision that the total number of headers, including each
%% individual object, not exceed six.  The \objectname{} macro, and its
%% alias \object{}, is used to mark each object.  The macro takes the object
%% name as its primary argument.  This name will appear in the paper
%% and serve as the link's anchor in the electronic edition if the name
%% is recognized by the data centers.  The macro also takes an optional
%% argument in parentheses in cases where the data center identification
%% differs from what is to be printed in the paper.

%\object{NGC 6624}, \objectname[M 15]{NGC 7078},
%\object[Cl 1938-341]{Terzan 8})}


\section{Introduction}

Sagittarius (Sgr) A*, the compact radio source at the Galactic center, is powered by accretion of a supermassive black hole of $\sim (3-4) \times 10^6 M_{\bigodot}$ \citep{S02, G04} in the prevailing stellar winds in the Galactic center region \citep{Melia92}. The bolometric luminosity of Sgr A* is more than 9 orders of magnitude lower than the corresponding Eddington luminosity, suggesting a radiatively inefficient accretion flow \citep{Melia01, Y03}. The radiative cooling processes therefore may be ignored while studying the dynamics of the accretion flow near the black hole, which simplifies the dynamical equations and the corresponding numerical algorithms significantly. 
Several general relativistic magneto-hydrodynamical (GRMHD) codes have been developed over the past few years to study the structure of non-radiative accretion flows near black holes quantitatively \citep{D03, Gam03, A05a}. Given the extensive observations available for this supermassive black hole, it provides perhaps the best opportunity to study the physical processes in the strong gravity near black holes with GRMHD simulations \citep{F00a, H08}.

The observed linear polarization and variability of the millimeter to near-infrared (NIR) emissions suggest that they are produced by the synchrotron process near the black hole \citep{A00, Melia00, G03, E04, E06}. The longer wavelength emissions are circularly polarized \citep{B01a} with weak linear polarization observed during frequent outbursts or flares \citep{Z04, Y07}, which in combination with the continuity of the centimeter to sub-millimeter spectrum suggests a synchrotron origin for the longer wavelength emissions as well \citep{F98, A05}. The analogy of Sgr A* with radio loud AGNs also favors a synchrotron scenario for the radio emission.
However, the longer wavelength flux densities are less variable than the millimeter and sub-millimeter flux densities, indicating emissions from relatively larger radii \citep{F00, Liu01, H04}. Indeed, the millimeter and higher frequency emissions have been explained reasonably well with an accretion torus within $\sim\ 10\ r_S$ of the black hole, where $r_S$ is the Schwarzschild radius \citep{Melia00, Gold05, Liu07, N07}; And high spatial resolution VLBI observations do show that the intrinsic size of the millimeter emission region is within $20\ r_S$ in radius \citep{B04, S05}. While the general relativistic (GR) effects are important for these shorter wavelength emissions originating near the black hole \citep{F00a, brom01}, the longer wavelength emissions are produced beyond $\sim10\ r_S$ so that the GR effects are insignificant and one may readily use the GRMHD simulations to model the observed emission without using the sophisticated GR ray-tracing radiation transfer codes for the polarized synchrotron emission \citep{Brod05, H08}.
%and variability of the  indicated that it is produced by relativistic charged particles via the synchrotron process near the black hole (Genzel et al. 2003; Eckart et al. 2004, 2006; Yusef-Zadeh et al. 2006). The longer wavelength radio emission is likely produced by the same process at larger radii \citep{Liu01}. 

High resolution X-ray observations with the {\it Chandra} space telescope shows that the quiescent X-ray excess from the direction of Sgr A* is extended and has ion emission lines in the spectrum, suggesting that the emission is mostly produced at large radii near the capture radius of the accretion flow by a plasma of a few keV in temperature \citep{B01, B03, X06}. X-ray flares are routinely observed from the direction of Sgr A* (Belanger et al. 2006). Their spectra can be fitted with a featureless power law \citep{Porquet03}, and they appear to be always accompanied by NIR flares, which are produced near the black hole \citep{E04, E06, Y06, Y08}. The correlation between the NIR and X-ray flares indicates a synchrotron self-Comptonization (SSC) origin for the X-ray flares \citep{Markoff01, Liu01}, though the X-ray flares may also be produced via the bremsstrahlung process enhanced by instabilities related to the accretion process \citep{L02, Tag06}. Recent observations of flares seem to favor a synchrotron process \citep{Liu01, D09}.

The quiescent state X-ray emission from the inner accretion flow therefore must be lower than the observed flux level. X-ray emission can be produced through both the SSC and bremsstrahlung processes. The observed high radio flux level and low upper limit of X-ray flux suggest that the radio emission cannot be produced via thermal synchrotron emission in a bounded flow \citep{Liu01}. It therefore has to be produced by unbounded outflows and/or by nonthermal populations of relativistic electrons.  Given the non-radiative nature of the GRMHD simulations, strong outflows and winds appear to be inevitable,
%GRMHD simulations naturally produce strong unbounded outflows, 
especially for highly spinning black holes \citep{N06, H07}. In this paper, we investigate whether emissions from these simulated accretion flows can count for observations of Sgr A*, especially the low frequency radio emissions, self-consistently.

 %\begin{figure}[bht] 
%\begin{center}
%\includegraphics[height=6.8cm]{observed.eps}
%\caption{
%The observed spectrum we must match using our simulations.  The spectrum in the radio and NIR frequencies must be matched by the synchrotron emission and the butterfly in the x-ray frequencies must not be exceeded by bremsstrahlung emission.}
%\end{center}
%\label{fig1.ps}
%\end{figure}
 %NIR, radio and x-ray emissions are observed quite frequently from the galactic center, this observed spectrum is shown on Figure 1. Our goal in this project is to simulate the physics within the accretion disk so that the observed spectrum can be properly predicted by the simulated model.  The simulation that we are trying to match to observation is the General Relativistic Magneto Hydro-Dynamical simulation, GRMHD. 

We carried out 2-dimensional (2D) GRMHD simulations with the Cosmos++ and HARM codes \citep{Gam03, A05a} assuming axis-symmetry of the accretion flow. The accretion disk therefore must be aligned with the equatorial planes of spinning black holes. %The simulation is two dimensional and the accretion disc is assumed to be axially symmetric about its pole.   The GRMHD simulation does not account for radiative cooling, this is a good assumption for the hot accretion disc in Sag A*.  
 The length scale is determined by the black hole mass. Since no radiative cooling is included, the density is scaled linearly with the mass accretion rate, and the gas temperature and plasma $\beta$, defined as the ratio of the gas pressure to the magnetic field pressure, are independent of the black hole mass and accretion rate. GRMHD simulations give the gas density, pressure, and magnetic field. The gas temperature can be derived from the equation of state. The mass of the supermassive black hole in Sgr A* is well-measured by observations of stellar orbits around the black hole \citep{S02, G04}. One therefore needs to adjust the accretion rate, black hole spin, and electron distribution to fit the observed spectrum of Sgr A*.
%This makes the particle density which is directly related to the magnetic pressure, our free parameter.  The goal is to use GRMHD simulations to match the observed spectrum while adjusting the particle density. 
We consider both a one temperature and a two-temperature model of electrons in the thermal synchrotron emission scenario.  For non-thermal synchrotron models, the electron distribution is assumed to follow a steep power-law with a low energy cutoff. %This paper focuses on modeling long wavelength radio emission produced at large radii. The GR effects, which affects the emission from within $10$ Schwarzschild radii of the black hole, therefore can be ignored.
%We ignore the regions in which the temperatures exceed physical values, the regions where general relativistic effects become important and the Doppler effect.  We do however account for the self-absorption of the accretion disk.  
In \S\ 2 we discuss the methods that we used to model the emission spectra from the simulated accretion flows.  A detailed discussion of our results and the implications are given in \S\ 3.

\section{Modeling the Emission Spectrum}

\begin{figure}[bht] 
\begin{center}
\includegraphics[width=5.3cm]{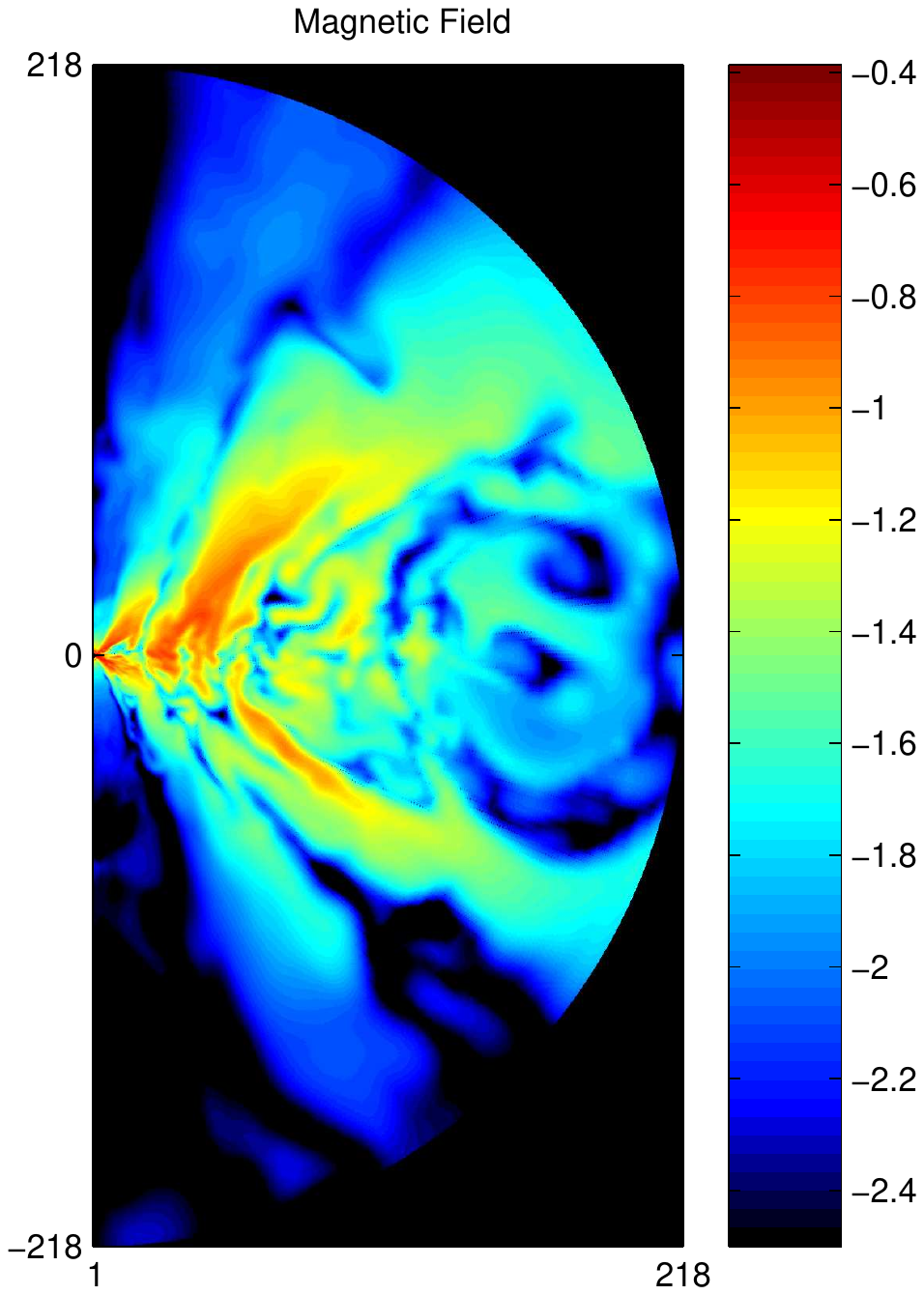}
%\hspace{0cm}
\includegraphics[width=5.3cm]{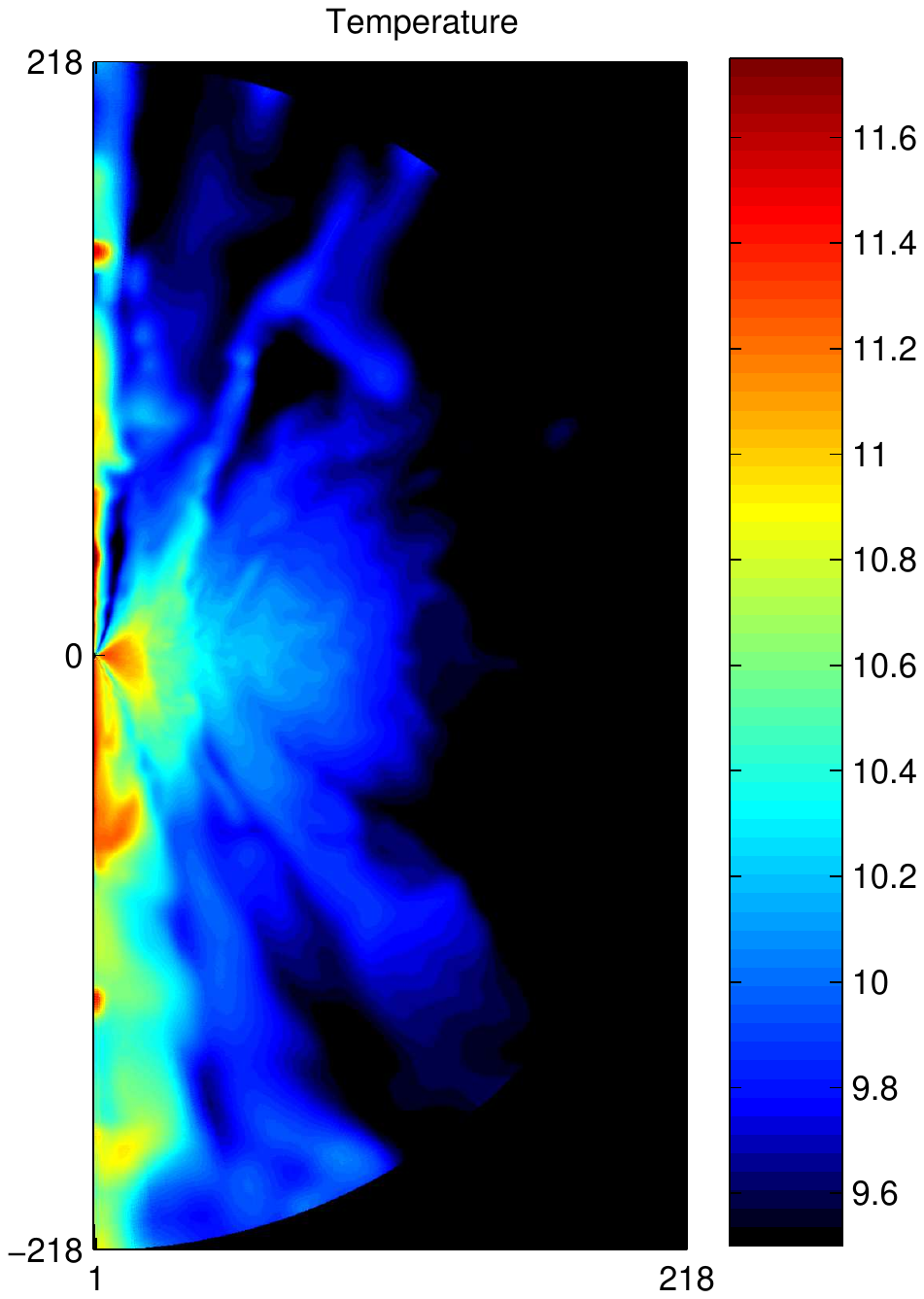}
\includegraphics[width=5.3cm]{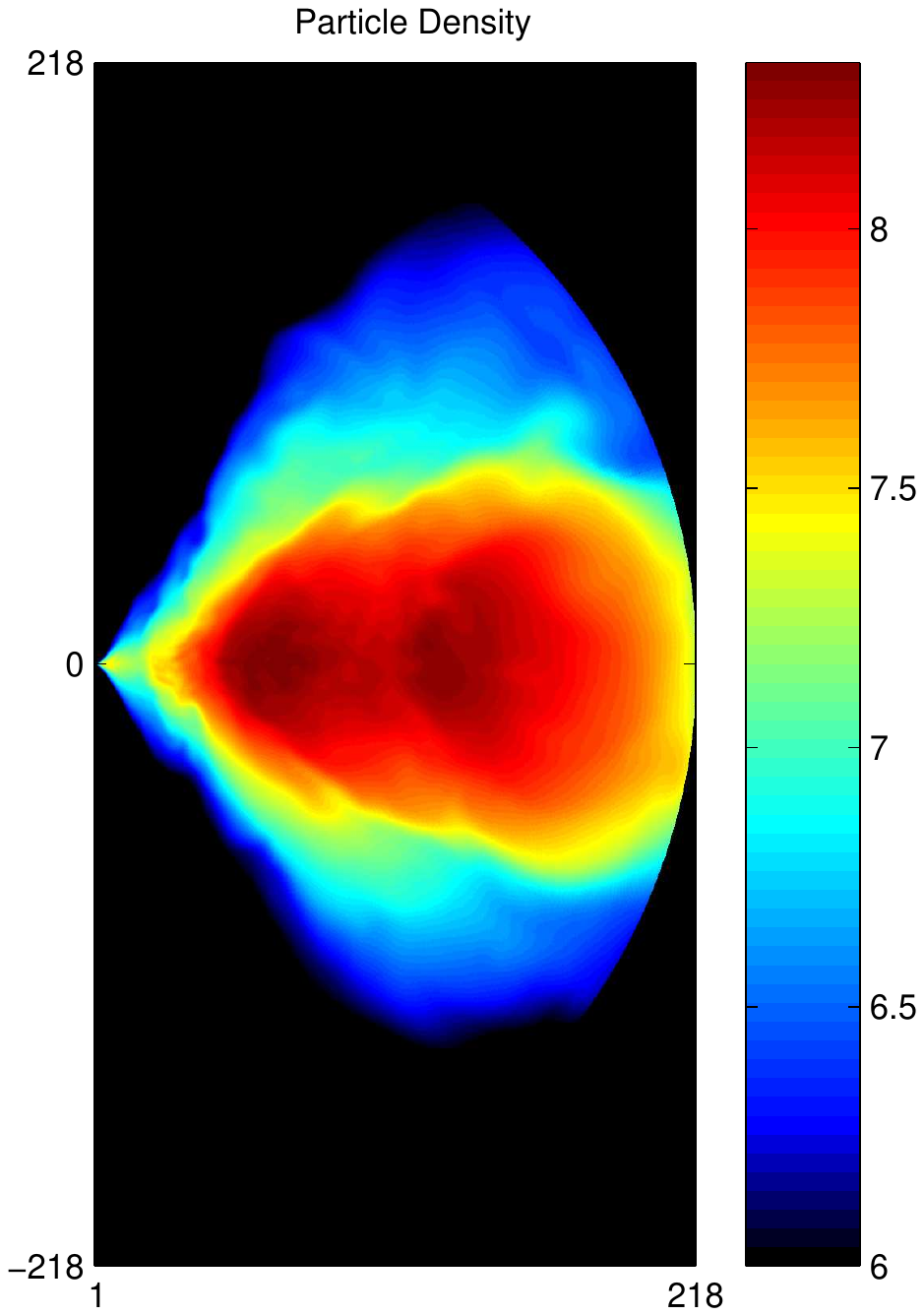}
%\hspace{0cm}
\end{center}
\caption{
{\it Left:} The magnetic field in units of Gauss in a logarithmic color scale for the simulation with the Cosmos++. 
{\it Middle:} Same as the left panel but for the temperature in Kelvin.
{\it Right:} Same as the left panel but for the particle number density in cm$^{-3}$. The outermost high density region corresponds to the position of the initial torus. If the simulation were to run longer this region of high density at about $100\ r_{S}$ would merge with the inner high density region.}
\label{fig1.ps}
\end{figure}

We run the Cosmos++ code over a simulation domain of $218\ r_S$ in radius. Initially, there is an equilibrium torus with the pressure maximum at $100\ r_S$. Under the influence of the strong gravity of the central black hole, the torus evolves and the magneto-rotational-instability sets in and induces the accretion process \citep{Balbus91}. 
%The first GRMHD simulation that we modeled was run on a short timescale and ran from 0-218 Schwarzschild radius, $r_S$.  
Specific attributes of the accretion disc were examined and the magnetic field, gas temperature, and particle density profiles are shown in Figure \ref{fig1.ps}. Because the simulation was run up to the dynamical time near $100\ r_S$, the initial torus structure has not relaxed completely and, beside the cusped  accretion torus in the inner region, which corresponds to the particle concentration within $100\ r_S$, there is another high density region at $\sim 100\ r_S$, which maps to the initial torus. If the simulation were allowed to run longer, this region of high density would be pulled closer to the black hole and merge with the inner accretion torus.  The simulation also produces extremely high temperature regions along the polar directions, the polar regions within a half open angle of $6^\circ$ and the inner $10\ r_S$ region are excluded in our modeling of the emission for this simulation. 

\begin{figure}[bht] 
\begin{center}
\includegraphics[width=5.3cm]{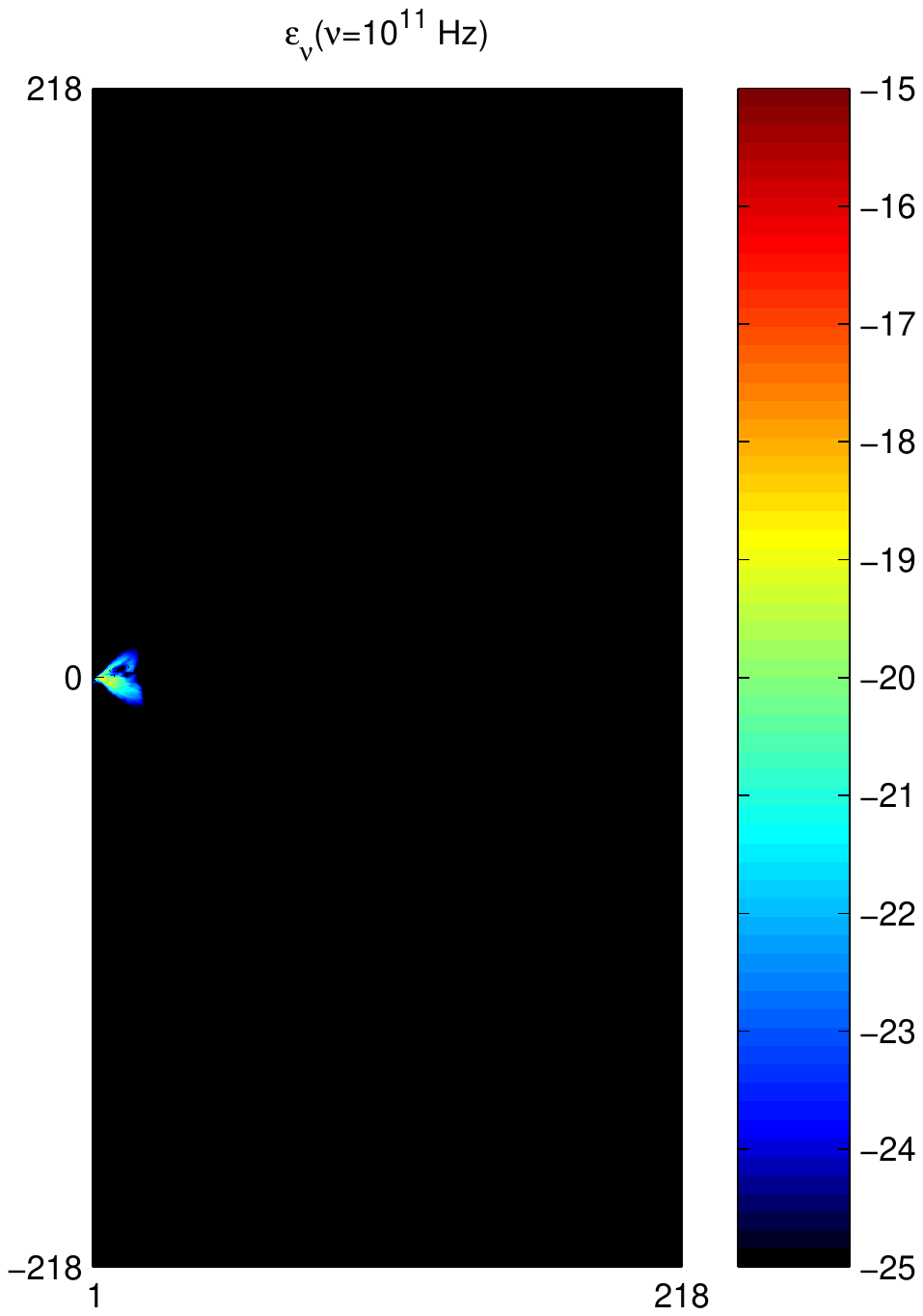}
\hspace{0cm}
\includegraphics[width=5.3cm]{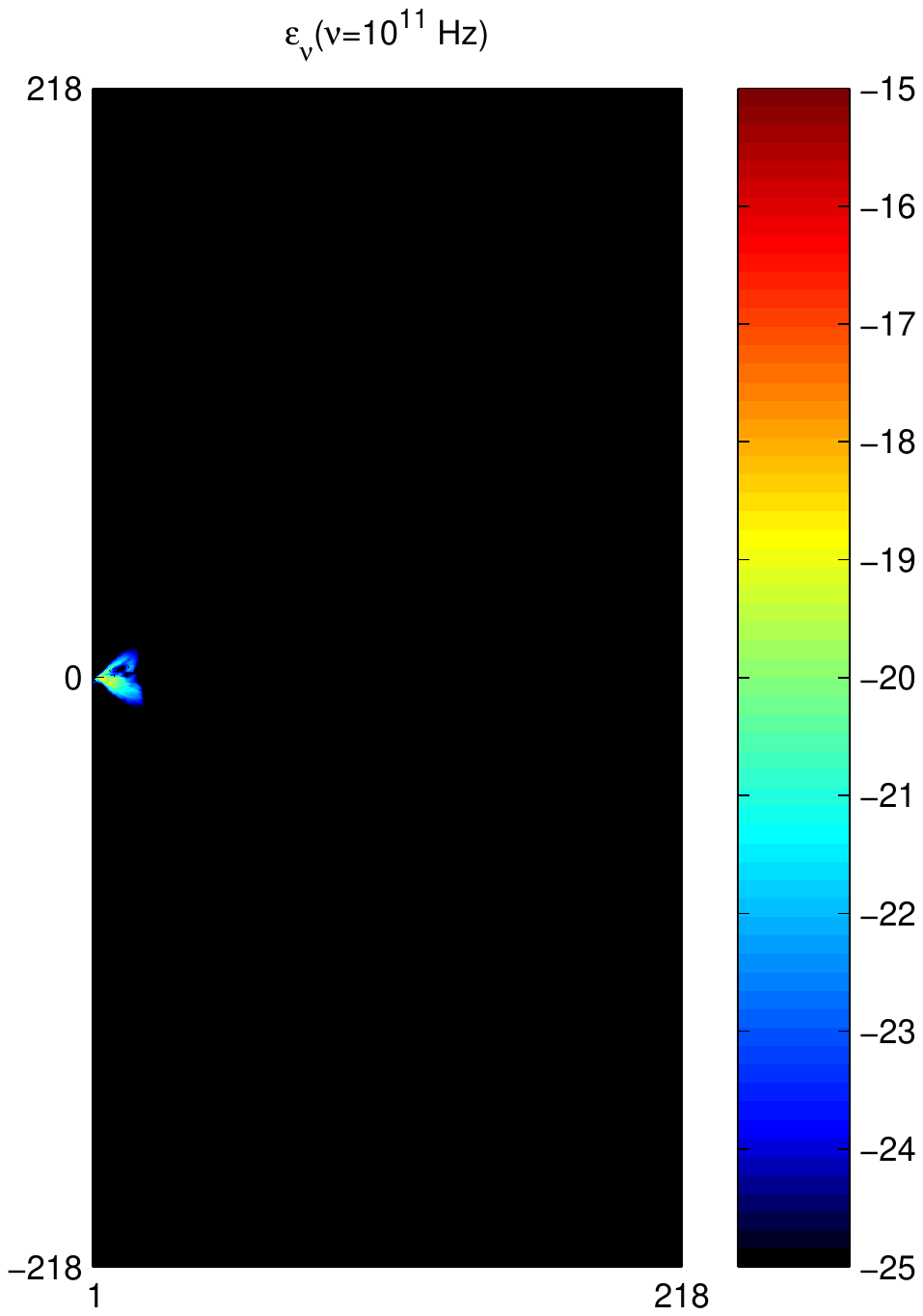}
\hspace{0cm}
\includegraphics[width=5.3cm]{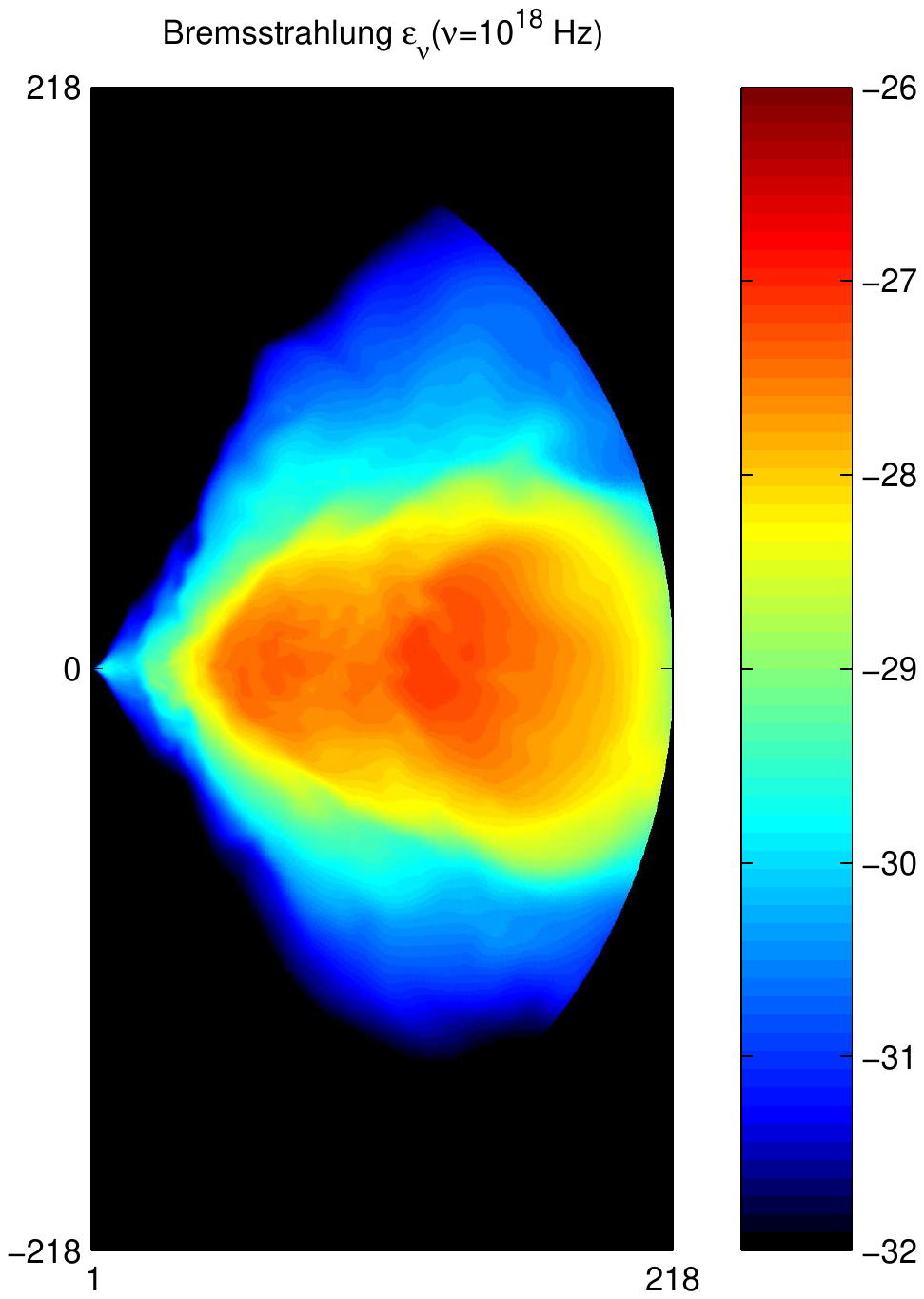}
\end{center}
\caption{
{\it Left:} The synchrotron emission coefficient in the plasma co-moving frame at $1.0$ GHz in the c.g.s. units.
{\it Middle:} Same as the left panel at $100$ GHz.
{\it Right:} The bremsstrahlung emission coefficient at $10^{18}$Hz, which corresponds to $\sim 4.1$ keV in energy.}
\label{fig2.ps}
\end{figure}

 We first consider the simplest case, where electrons and ions reach thermal equilibrium throughout the simulation domain, so that the electron temperature is given by the gas temperature.
%The synchrotron emission is produced by thermal electrons. 
The synchrotron emission coefficient at frequency $\nu$ is
given by \citep{Pac70, Maha96, Liu06}
\begin{equation}
{\cal E}_\nu(\nu)={\sqrt{3} e^3\over 4\pi m_e c^2} B\,n\,x_{m}\,I(x_{m}) 
\,,
\end{equation}
%This is an incomplete picture.  Equation (2) does not account for the self absorption of the gas, where emitted light will be absorbed by the accretion disc and re-emitted.  The flux intensity $I_{\nu}$ has to be corrected by 
where
\begin{eqnarray}
x_{m}&=&{\nu\over \nu_c}\equiv {4\pi m_e c\ \nu\over 3eB\gamma_{c}^2}\,,
\\
I(x_{m})&=&4.0505 x_{m}^{-1/6}(1+0.40 x_{m}^{-1/4}+0.5316 x_{m}^{-1/2})\times \exp(-1.8899 x_{m}^{1/3})\,,
\\
\gamma_{c}&=&{k_{\rm B} T_{e}\over m_{e} c^2}
\,,
\end{eqnarray}
where $T_e=T=P_{gas}m_p/2\rho k_{\rm B}$ is the electron temperature and  $T$, $P_{gas}$, $\rho$,  $B$, $n$, $e$, $m_e$, $c$, and $k_{\rm B}$ indicate the gas temperature, pressure, mass density, magnetic field, particle density, the elementary charge unit, electron mass, speed of light, and the Boltzmann constant, respectively. 
The left and middle panels of Figure \ref{fig2.ps} show the profiles of ${\cal E}$ at $1.0$ and $100$ GHz, respectively.
The corresponding absorption coefficient $\kappa_{\nu}$ and the opacity $\tau$ along the light of sight of the
accretion disk is given, respectively, by 
\begin{equation}
\kappa_{\nu}={{\cal E}_\nu\over 2 \gamma_{c} m_{e} {\nu}^2}\,, \ \ \ \
\tau=\int {\kappa_{\nu}}\, dl
\,,
\end{equation}
where the integration of $\tau$ is along the light of sight from the observer into the source region.

\begin{figure}[bht] 
\begin{center}
\includegraphics[width=5.5cm]{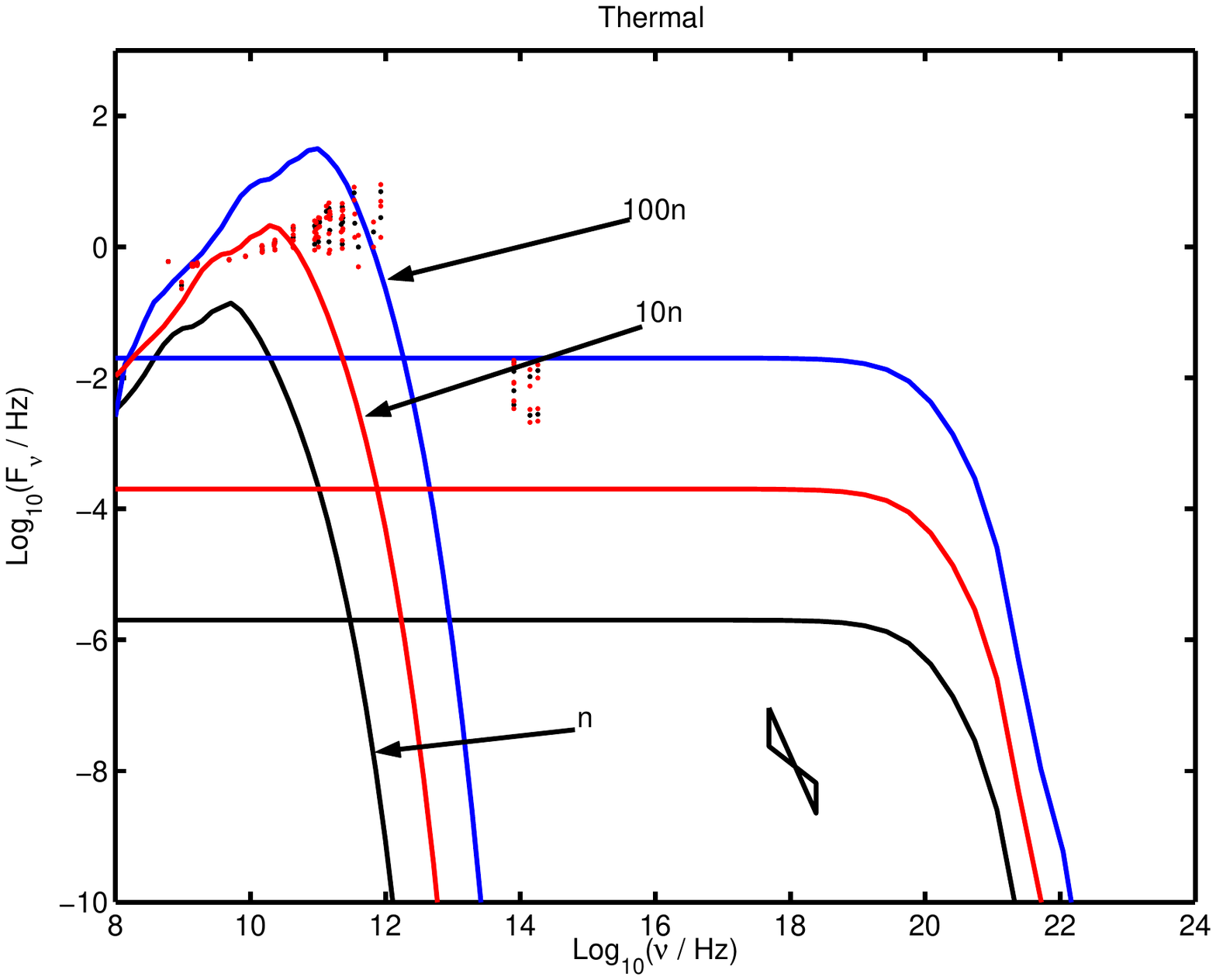}
\hspace{-0.0cm}
\includegraphics[width=5.5cm]{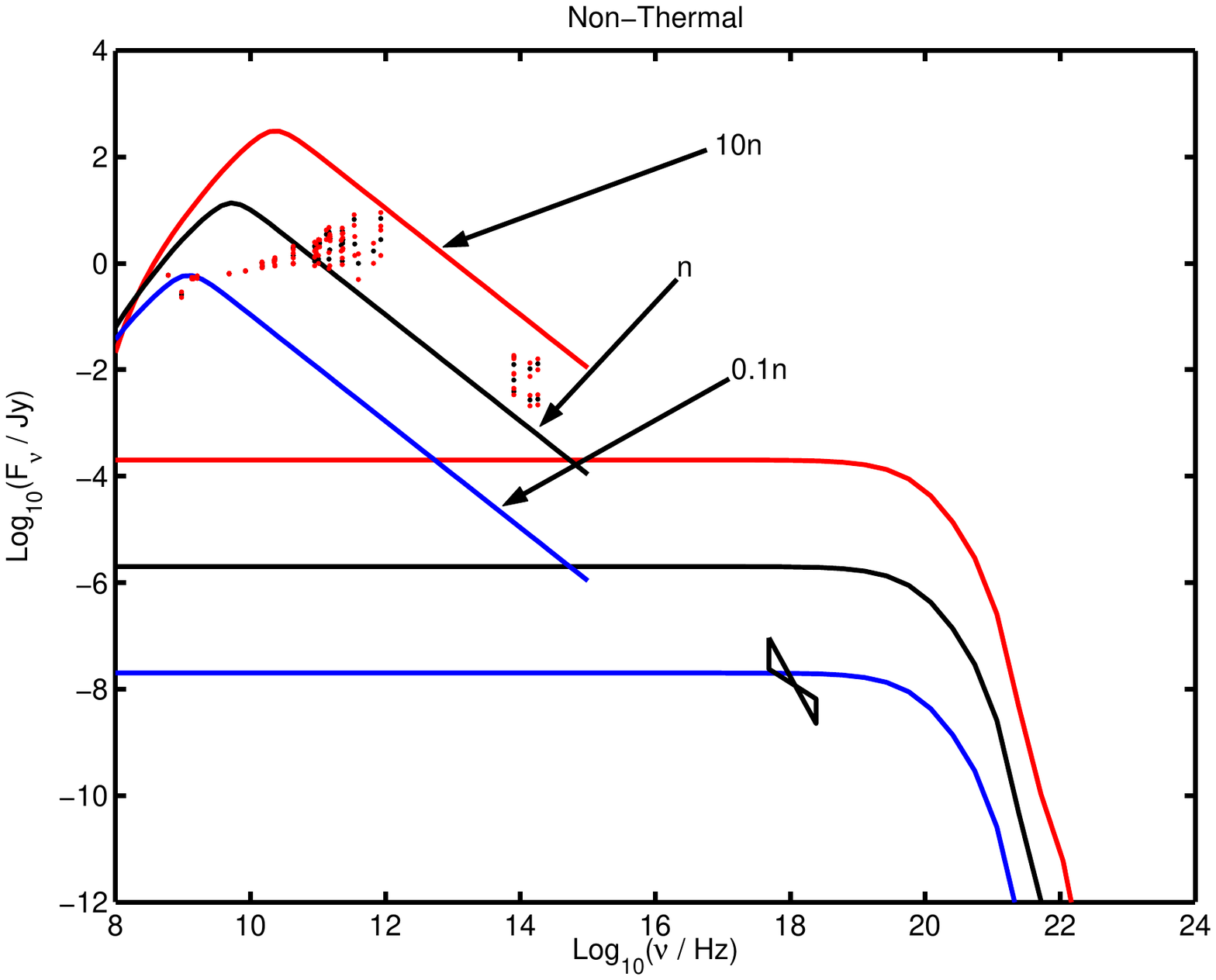}
\end{center}
\caption{
{\it Left:} Thermal synchrotron and bremsstrahlung emission spectra for the one temperature model.  The black, red, and blue lines correspond to the particle density in Figure \ref{fig1.ps}, 10, and 100 times this density, respectively.  As expected, the bremsstrahlung flux level scales with the density squared. The radio to sub-millimeter data was obtained from observations over the past decade. The scattering of the millimeter and sub-millimeter data is caused by the variability of the source. The NIR flux densities are for flares observed during the past few years and therefore should be considered as upper limit for the ``quiescent'' state emission from the accretion flow. The butterfly is given by {\it Chandra} observations of the quiescent state X-ray emission, which appears to be dominated by thermal emission at large radii. Thus it also should be interpreted as an upper limit for X-ray emission from the small scale accretion flow. All the bremsstrahlung spectra exceed this upper limit for X-ray emission.
{\it Right:} Same as the left panel but for the non-thermal model with a single power-law electron distribution with a low energy cutoff. The synchrotron spectra are showed up to $10^{15}$ Hz.  The density normalization is indicated in the figure. Compare with the thermal models in the left panel, the nonthermal model is more efficient in the synchrotron emission process. However, to match the observed radio flux level, the bremsstrahlung X-ray flux still exceeds the observed upper limit.}
\label{fig3.ps}
\end{figure}

Then the observed flux density from the accretion flow \citep{Melia01}
\begin{equation}
F_{\nu}(\nu)={1\over D^2}\int I_{\nu}\, ds\,,
\end{equation}
where
$$ I_{\nu}=\int {\cal E}_{\nu} \exp(-\tau)\, dl\,, $$
is the specific intensity and
$D$ is the distance to the Galactic Center and the integration of $F_\nu$ is over the projected area of the source.
For a face-on disk with the observer located at $z=D$, the radio spectra are shown in the left panel of Figure \ref{fig3.ps} for several values of the density normalization. These values are chosen so that the model spectra embrace the observed radio flux densities. The red line with a 10 times higher density profile than that shown in Figure \ref{fig1.ps} gives the best fit to the observed spectrum.
%We needed to also calculate the emission in the x-ray spectrum.  We receive observations of x-ray emissions on a regular basis and our model of the galactic center should match these observations if it is to be complete.  X-rays are believed to result from Bremsstrahlung emission, the  radiation from deceleration of charged particles after collisions with other charged particles.
%We also consider the Bremsstrahlung emission so that we have a control on our data.  While trying to match the observed spectrum we could mistakingly create an accretion disc that has non-physical properties, by calculating the emission in the x-ray using a different method and checking it against observed data we can assure that if we are under the limit imposed on the bremsstrahlung emission then the model we have for the synchrotron emission is correct.  

However, the high gas density implies strong X-rays via the bremsstrahlung process from the two torii evident in the density profile of Figure \ref{fig1.ps}.
The bremsstrahlung emission coefficient is calculated by \citep{RL79}: 
\begin{equation}
{\cal E}_\nu=8.5\times 10^{-39} n^2 T_e^{-1/2} \exp(-h \nu / k_{\rm B} T_e)
\,,
\label{brem}
\end{equation}
where $h$ is the Planck constant. The right panel of Figure \ref{fig2.ps} shows the bremsstrahlung emission coefficient at $\sim 4.1$ keV and the corresponding emission spectra are also indicated in the left panel of Figure \ref{fig3.ps}. The bremsstrahlung emission exceeds the observed upper limit for all the three density normalizations. This simple model therefore cannot produce the observed radio flux level without exceeding the X-ray upper limit.
%Bremsstrahlung emission does not require the self-absorption term that synchrotron emission does given by Equation (3), we are mostly concerned with the emission in the x-ray
%spectrum.   Self absorption is inversely proportional to the square of the frequency, the radiation with which we are concerned has virtually no self-absorption in the x-ray spectrum.
%We modeled the first simulation using the thermal synchrotron and Bremsstrahlung emission, Figure 4 shows the results.

%As we increased the particle density to try and match the observed spectrum we exceeded the upper limit of the x-ray emission produced by Bremsstrahlung emission.  

It is well-known that nonthermal synchrotron emission is more efficient than thermal synchrotron. We next study whether a simple nonthermal model with a single power-law electron distribution can account for the radio flux without violating the X-ray upper limit.
%needed a model that was more effective at producing synchrotron emission in the NIR and radio frequencies.  We examined the non-thermal synchrotron emission. 
%The non-thermal synchrotron emission model is much more effective at producing emission in the NIR and radio frequencies, while hopefully keeping the x-ray emissions from Bremsstrahlung emissions relatively low.  
%This non-thermal  model assumes that the distribution of electrons can be given by a power law:
The electron distribution is given by
\begin{equation}
N(E)=N_0 E^{-p}\ \ \ \ {\rm for} \ \ \  E>E_0\,,
\end{equation}
where $E_0$ is the low energy cutoff.  $N_0$ and $E_0$ can be calculated with
\begin{eqnarray}
\displaystyle\int^\infty_{E_0} N_0 E^{-p}\,dE = n\,,
\\
\displaystyle\int^\infty_{E_0} N_0 E^{-p +1}\,dE = 3nk_{\rm B} T_e
\,.
\end{eqnarray}
%the above two integrals are solved to obtain values for $E_0$ and $ N_0$ we assumed that $E_{\infty} = E_0 \times 10^5$.  
%at $\gamma=6$ the non-thermal model and the thermal model were essentially the same.  
The corresponding emission and absorption coefficients are given respectively by  \citep{Pac70}
\begin{eqnarray}
{\cal E}_\nu=c_5(p) N_0 [B \sin\alpha]^{(p +1)/2} \left({\nu\over 2 c_1}\right)^{(p -1)/2}\,,
\\
\kappa_{\nu}=c_6(p) N_0 [B \sin\alpha]^{(p+2)/2} \left({\nu\over 2 c_1}\right)^{-(p+4)/2}
\,,
\end{eqnarray}
where $\alpha$ is the angle between the magnetic field and the line of sight and $c_i$ are defined in \citet{Pac70}. In our calculations we chose $p=3$ to avoid significant NIR emission in the quiescent-state. As the $p$ increases, the  non-thermal emission model became less efficient and closer to the thermal model. 
% the $dl$ that the opacity and emission intensity in equation (3) are integrated along, and While Equations (2) and (3) were still used to calculate the flux, emission intensity and the opacity of the accretion disc.  
The right panel of Figure \ref{fig3.ps} show the spectra for several density normalizations, where the synchrotron spectrum is cut off artificially at $10^{15}$Hz and the bremsstrahlung spectra is given by the thermal formula of equation (\ref{brem}). Compared with the thermal model, more radio emission is produced for a given density normalization. However, to produce the observed radio flux level, the bremsstrahlung emission still exceeds the observed upper limit.
 
%Figure 4 shows the results for the first simulation using the non-thermal model.  Unfortunately the outcome was the same, see Figure 4, by trying to correctly match the observed spectrum in the NIR and radio frequencies we surpassed the upper limit imposed by Chandra observations. 

\begin{figure}[bht] 
\begin{center}
\includegraphics[width=5.3cm]{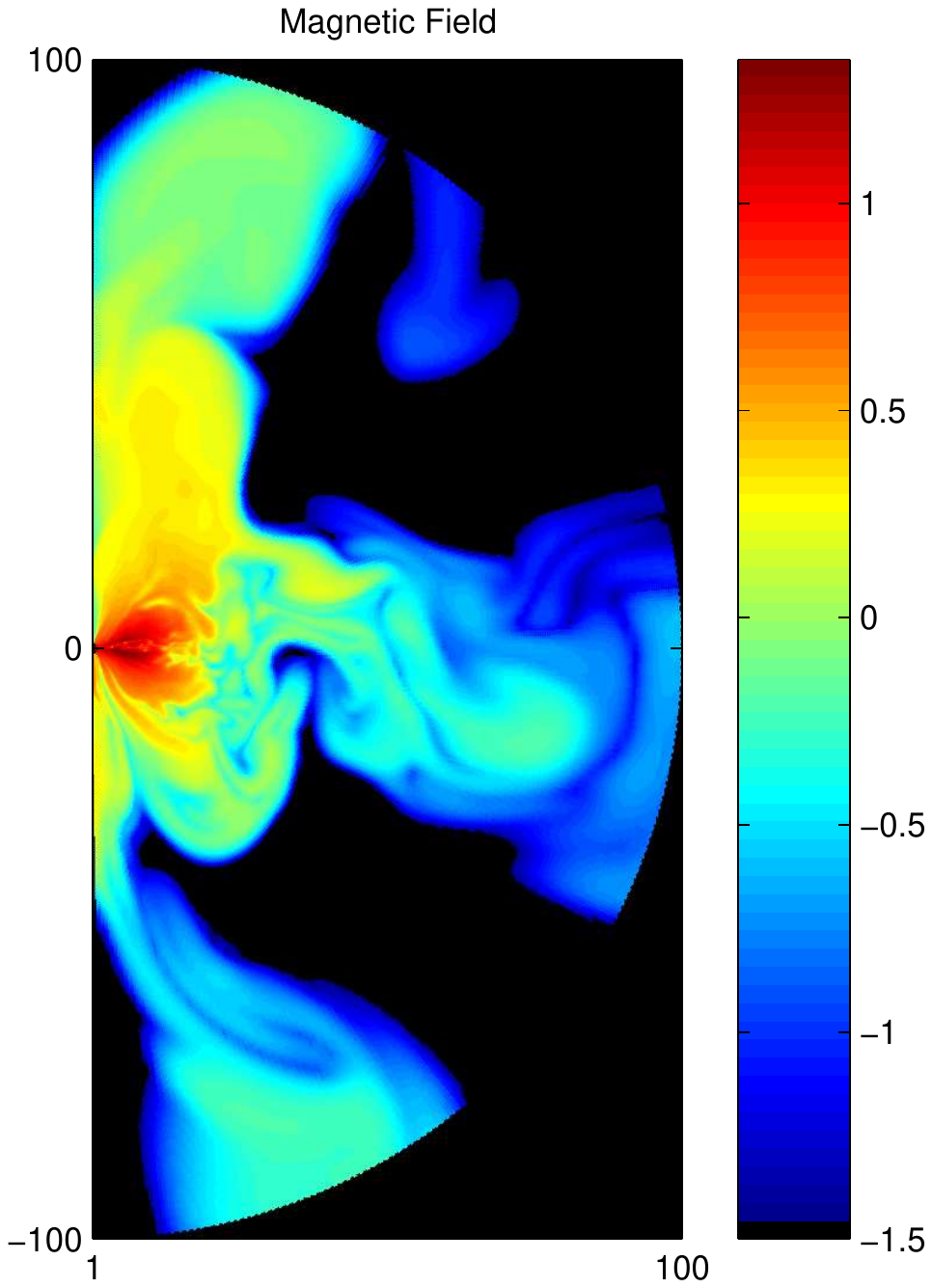}
\hspace{0cm}
\includegraphics[width=5.3cm]{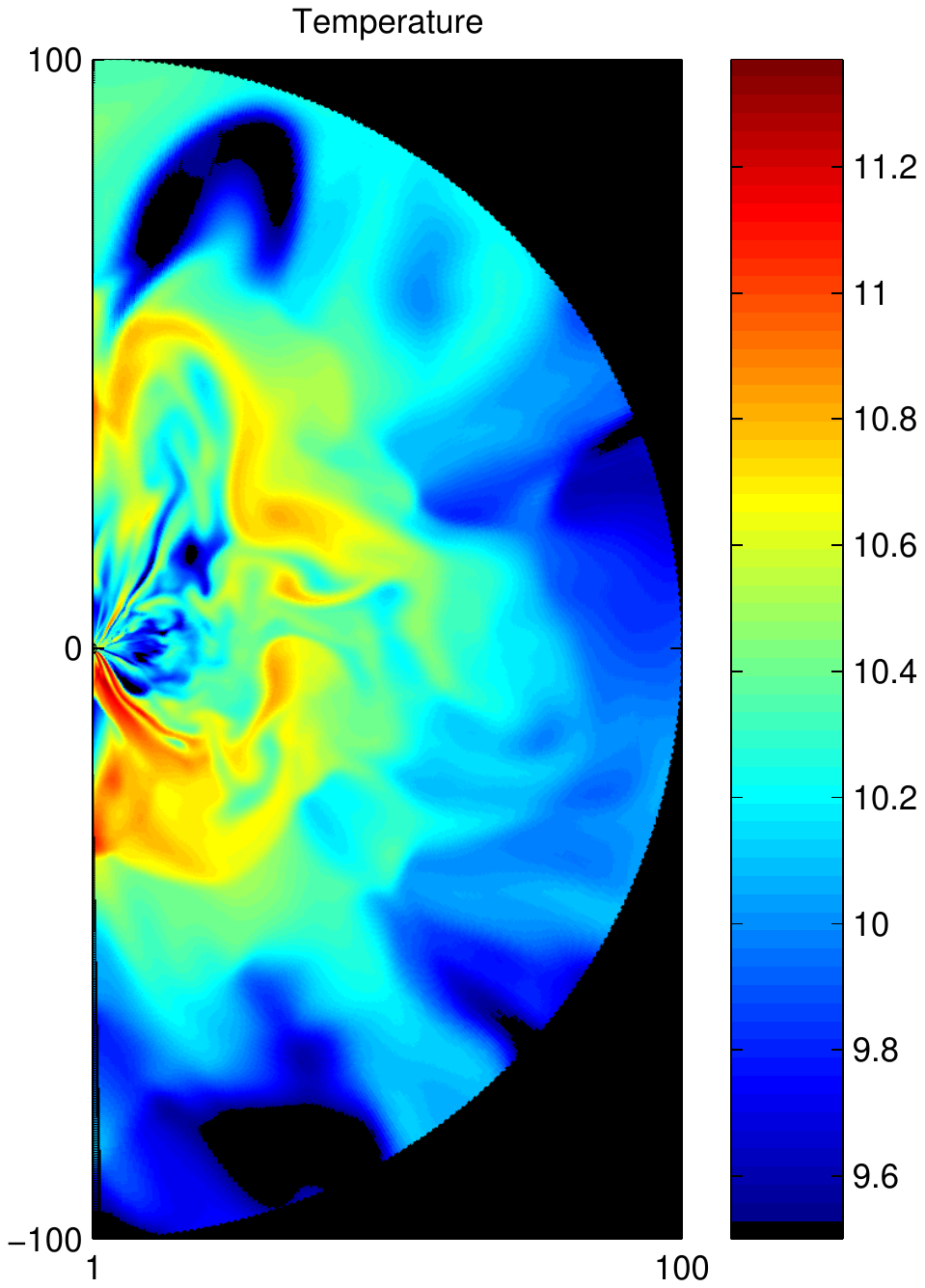}
\hspace{0cm}
\includegraphics[width=5.3cm]{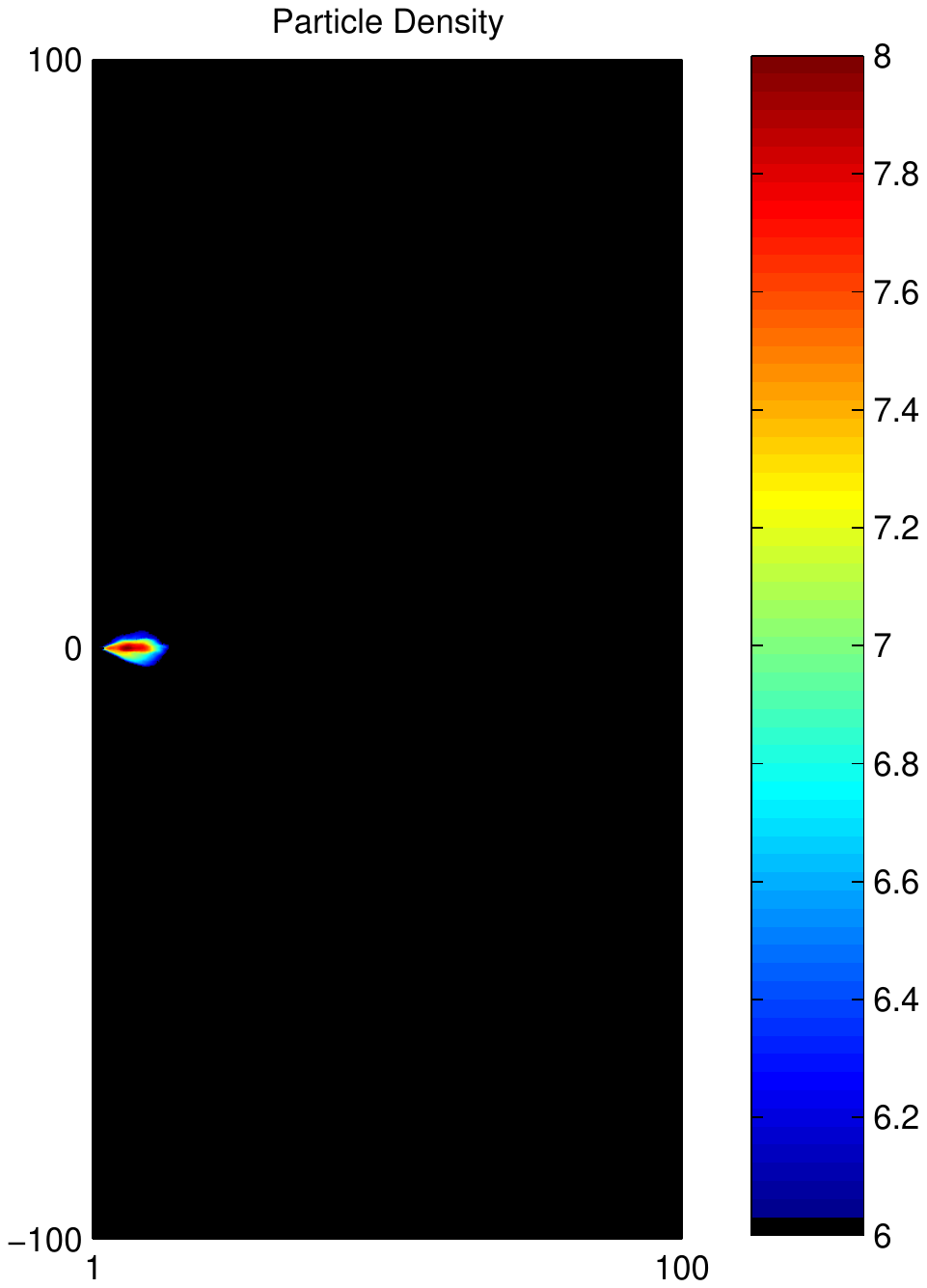}
\end{center}
\caption{
Same as Figure \ref{fig1.ps} but for a simulation with the HARM with a zero black hole spin.}
\label{fig4.ps}
\end{figure}

While the outer torus of the simulation is expected to merge with the inner torus if the simulation is run for a long enough time, the inner torus is already relaxed at the present epoch of the simulation. Therefore a longer run may remove the bremsstrahlung emission from the outer torus, we don't expect the overall bremsstrahlung X-ray flux to decrease by more than a factor of 2. It is therefore challenging to produce the observed radio flux level from Sgr A* without exceeding the X-ray upper limit with this simulation.
The high X-ray flux as compared with the radio flux is a direct consequence of the high values of plasma $\beta$ given by this simulation.
\begin{equation}
\beta ={ P_{gas}\over B^2/8\pi}\,.
\end{equation}
For this simulation, $\beta$ is always greater than $100$ except in the polar regions, where the GRMHD simulation is not reliable and we have excluded these regions while obtaining the emission spectra. The bremsstrahlung emission is determined by the gas temperature and density. The synchrotron emission also depends on the magnetic field. For high values of $\beta$, the magnetic field is weak. It is therefore difficult to produce significant radio emission without making X-rays through the bremsstrahlung. Simulations that are capable of producing lower values of $\beta$ are needed to overcome this challenge.
%the beta factor is a ratio of gas pressure to magnetic pressure. 
%We needed another simulation that decreased the accretion disc $\beta$ factor, i.e increased the magnetic field relative to the particle density.   
We also noticed that the radio spectra of both the thermal and non-thermal model are too hard to fit the observed spectrum. Decreasing the value of plasma $\beta$ alone may not explain the observations.
% i.e.the observed spectrum is relatively flat and peaks at a different place then either of our current calculations of the first simulation.  We needed a way to flatten the model and increase the frequency where the model's spectrum peaks.   

\begin{figure}[bht] 
\begin{center}
\includegraphics[width=5.2cm]{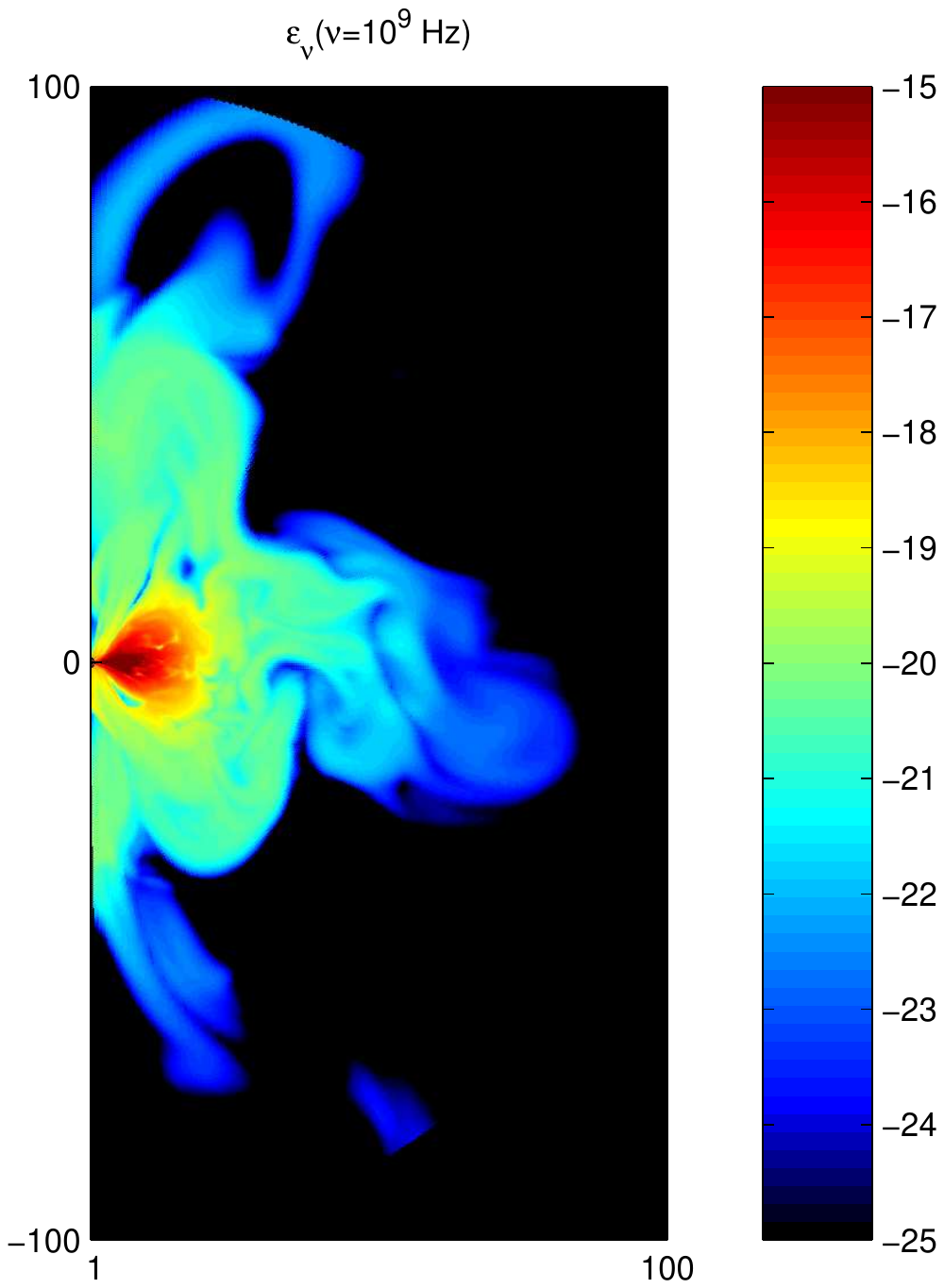}
\hspace{0cm}
\includegraphics[width=5.1cm]{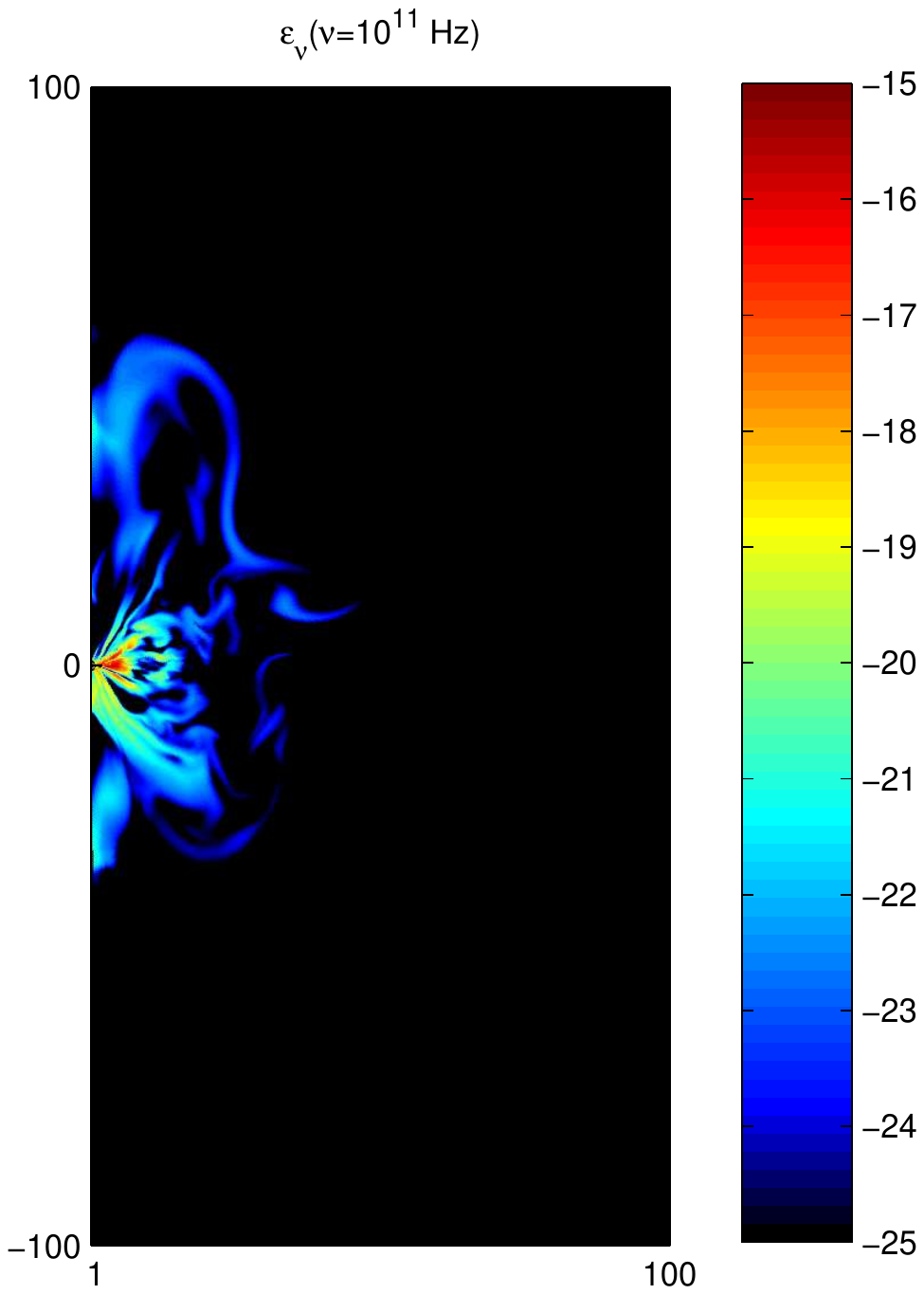}
\hspace{0cm}
\includegraphics[width=5.0cm]{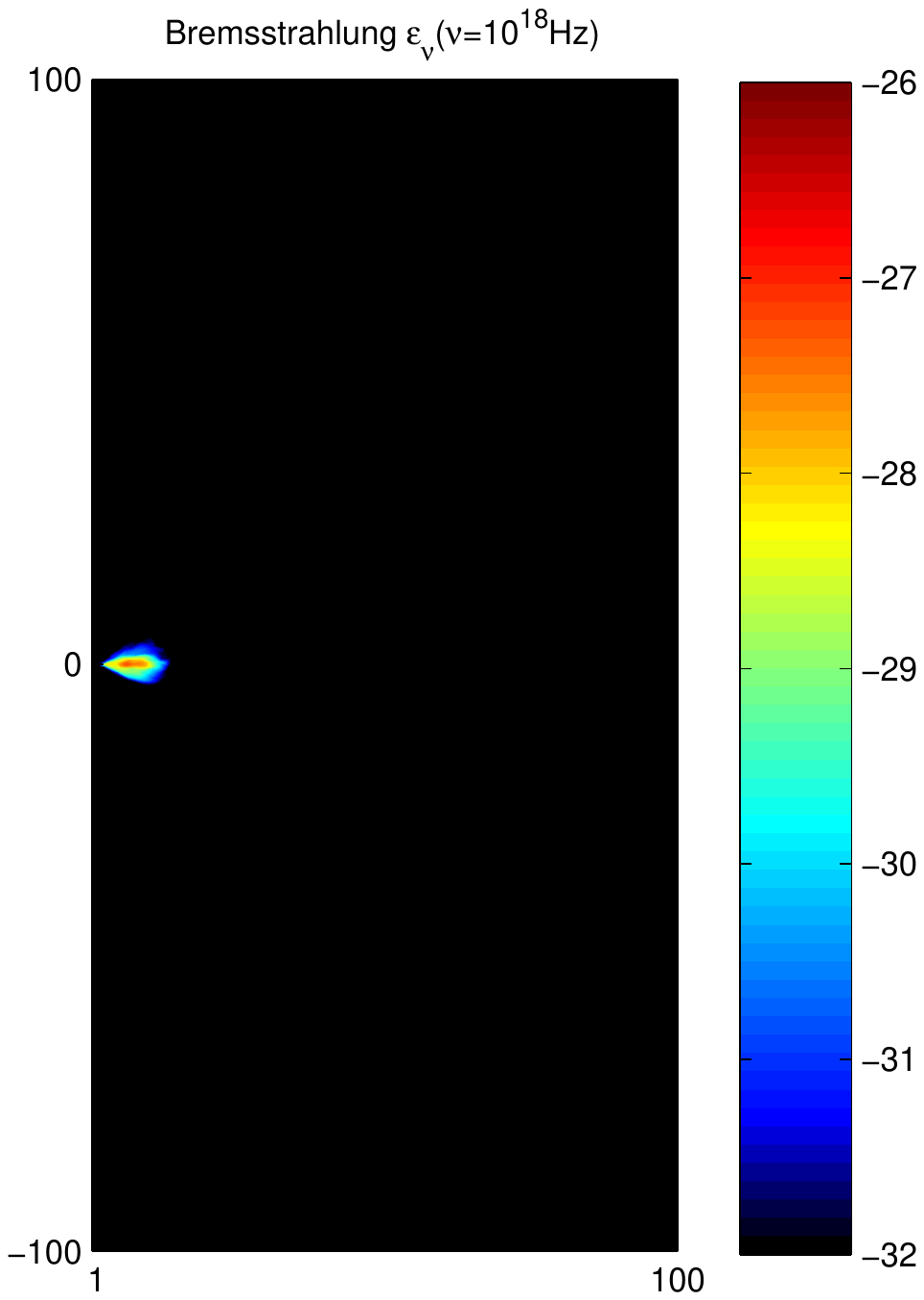}
\end{center}
\caption{
Same as Figure \ref{fig2.ps} but for the simulation of Figure \ref{fig4.ps}.
}
\label{fig5.ps}
\end{figure}

\begin{figure}[bht] 
\begin{center}
\includegraphics[width=5.5cm]{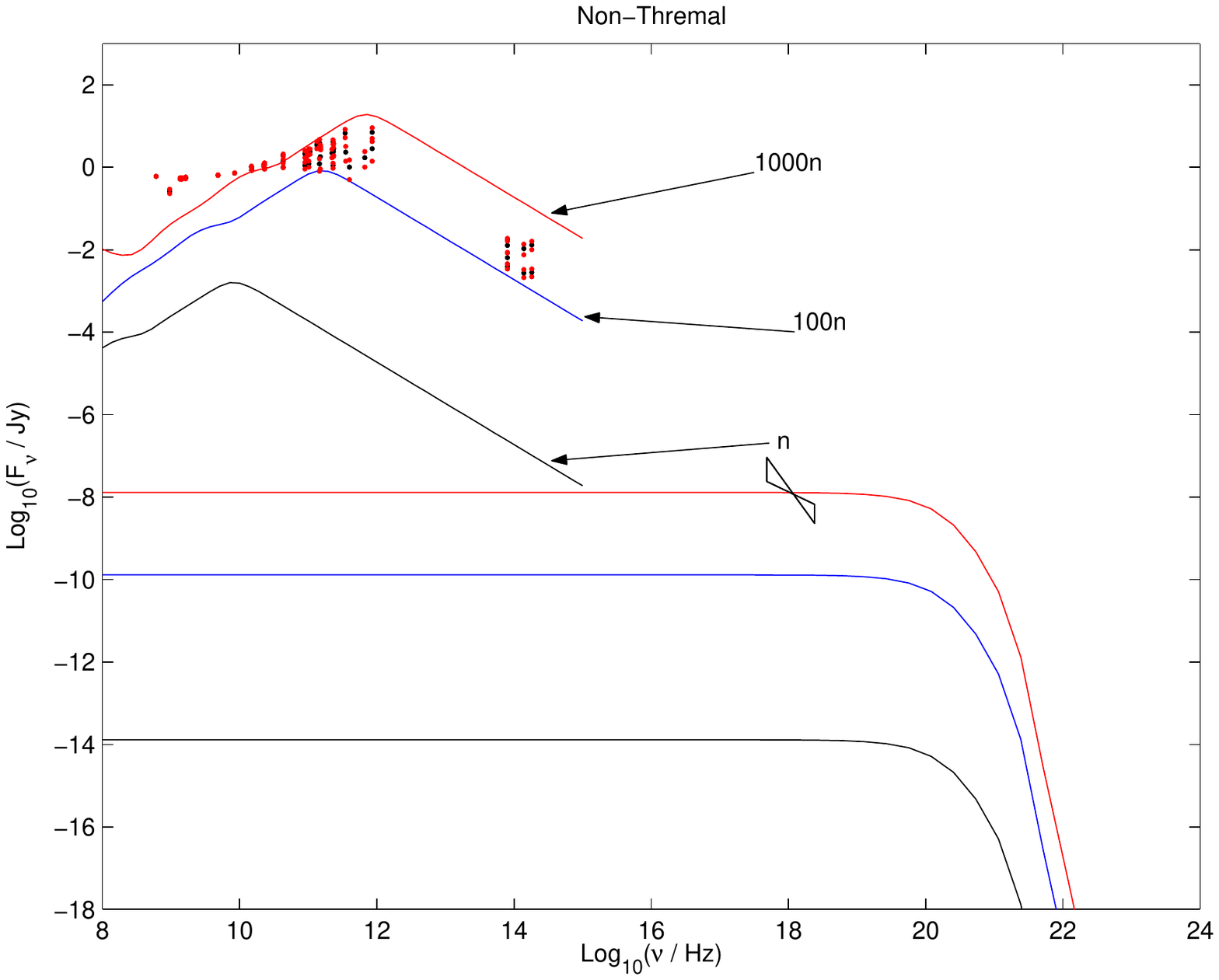}
\hspace{-0.0cm}
\includegraphics[width=5.5cm]{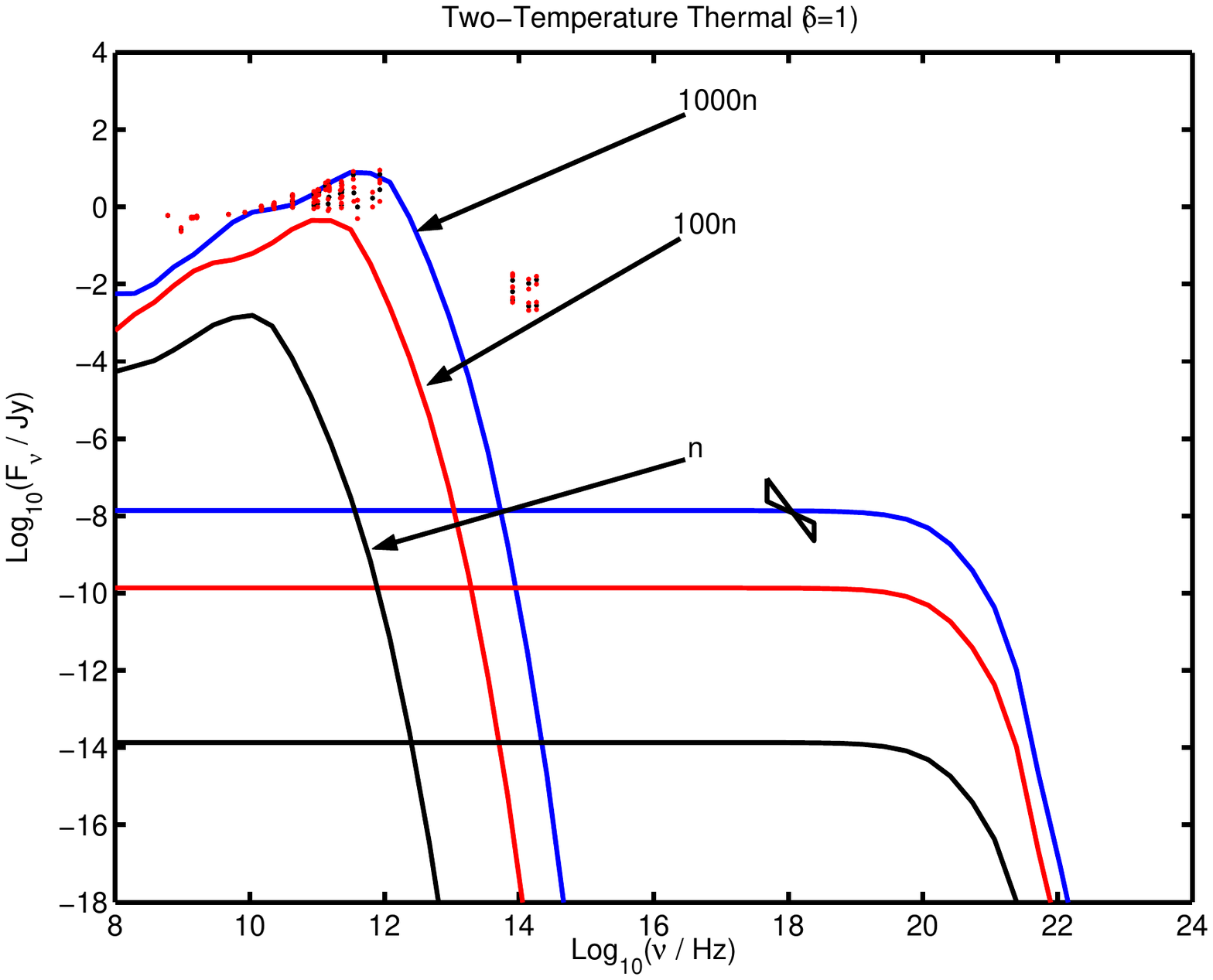}
\end{center}
\caption{
{\it Left:} The same as the left panel of Figure \ref{fig3.ps} but for the simulation of Figure \ref{fig4.ps}.
{\it Right:} The same as the left panel except that electron temperature is lower than the gas temperature toward small radii as given by equation (\ref{tt}) with $R=10\ r_S$ and $\delta=1.0$.}
\label{fig6.ps}
\end{figure}

Errors in the divergence of B are handled differently in the Cosmos++ than the HARM code \citep{Gam03, A05a}. The latter uses a contrained transport schemes for evolving the induction equation whereas Cosmos++ does not. As a consequence, the HARM code appears to be able to produce lower values of plasma $\beta$. We ran the HARM code in a domain of $100\ r_S$ in radius with the torus pressure maximum initially at $\sim 6\ r_S$. The simulated disk structure in a late time when the initial torus has completely relaxed is shown in Figure \ref{fig4.ps}. The simulation produces more than 10 times stronger magnetic fields with comparable values for the density maximum. Figure \ref{fig5.ps} gives the corresponding emission coefficient maps. Although the synchrotron emission may be comparable to that in Figure \ref{fig4.ps}, we expect much less bremsstrahlung X-rays due to the compactness of the torus in this latter simulation. 
%The second simulation that we examined was allowed to run longer then the first, there was no longer the outer region of high density.  The second simulation also covered a smaller range only $0-100 r_s$, this is where we suspect that most  of the emission comes from anyway.
The left panel of Figure \ref{fig6.ps} shows the spectra of the thermal model with several density normalizations. The bremsstrahlung X-ray emission is far below the observed upper limit. However, the radio spectra are still not flat enough to fit the observed spectrum. 

Higher frequency emission is produced at relatively smaller radii, where the magnetic field is strong and the gas temperature is high. The fact that the model predicted spectrum is harder than the observed spectrum suggests that emission from small radii needs to be suppressed to produce a softer spectrum. The synchrotron cooling time of electrons at smaller radii is shorter. It is possible that the electron temperature is actually lower than the gas temperature at small radii. Modeling of the emission from a small accretion torus also favors a lower electron temperature than the gas temperature \citep{Liu07, N07, H08}.
 %To soften up the spectrum, flatten it out, 
We model the electron temperature with 
\begin{equation}
T_e=\left({r\over R}\right)^{\delta} T
\,,
\label{tt}
\end{equation}
where $R$ is the radius, beyond which electrons reach thermal equilibrium with the ions and $\delta$ is a free parameter determined by the electron heating and cooling processes.
%the  ratio of the radius at which the temperature is defined to the largest radius of the accretion disc model.  We varied  $\delta$ in order to best match the  observed spectrum. 
%used a model that varied the electron temperature as a function of radius from the black hole. Using this two temperature model we found that we  reduced the overall emission flux across the spectrum.  In doing this we were able to flatten out the peak of the emission flux spectrum to more closely match the observed data.  The two temperature model that we used was given by:

%Since the non-thermal model did not rely on temperature we had to change the way in which we calculated $N_0$ equation (13) became:
 %\begin{equation}
%\displaystyle\int^{E_{\infty}}_{E_0} N_0 E^{-\gamma+1}\,dE = \epsilon_0 ({r\over R})^{\delta}\,,
%\end{equation}

%  \begin{figure}[bht] 
%\begin{center}
%\includegraphics[width=7.4cm]{sim22T.eps}
%\hspace{-0.0cm}
%\includegraphics[width=7.4cm]{sim22nT2.eps}
%\end{center}
%\caption{
%{\it Left:} The second simulation two-temperature thermal synchrotron emission.  %The spectra is given for $10^6$ times, $10^4$ times and $10^5$ times the %original simulation particle density. Although the two temperature model with %$\delta =.4$ more closely matches the  observed radio and NIR spectrum the %upper limit imposed on the bremsstrahlung emission is exceeded. 
%{\it Right:} The second simulation non-thermal synchrotron emission.  We %encounter a similar problem with exceeding the upper bound on the X-ray %emission, even though the shape of the synchrotron emission looks promising %with $\delta=1$.
%}
%\label{fig5.ps}
%\end{figure}

There was no need to apply the two-temperature model to the first simulation because it would decrease the amount of NIR and radio emissions without decreasing the bremsstrahlung emission values, we would still have the same problem of exceeding the upper limits imposed by {\it Chandra} on the X-ray emission.  We applied this two-temperature model with $R=10\ r_S$ and $\delta =1.0$ to the second simulation and calculated the thermal emission spectrum.  The right panel of Figure \ref{fig6.ps} shows the results.  The model with a 1000 times higher density than that in Figure \ref{fig4.ps} fits the radio spectrum reasonably well. The underproduction of radio emission below 10 GHz is due to the relatively small simulation domain of $100$ $r_S$. Most of the longer wavelength emission is produced beyond $100\ r_S$.  However, the model predicted X-ray flux level just matches the X-ray upper limit. So this model fits the observations marginally. Including the Doppler effect and considering the possibility of nonthermal emission may improve the fitting.

\section{Conclusions}

Sgr A* is a radio source powered by accretion into a black hole.  Modeling the emission from this radio source at the Galactic center provides us with knowledge of the dynamics of the accretion flow near black holes and helps us further understand the plasma physics processes of the accretion torus.  We were unsuccessful in our attempt to reproduce the observed spectrum using simulations with very high values of plasma $\beta$. 
%With what we learned in this project and the ground work that we laid it should not be too hard to figure out what is missing from the current models.   
With the HARM code, lower values of plasma $\beta$ are produced. Although we were able to apply a two temperature model to this simulation to correctly match the shape of the observed radio emission spectrum, we had the reoccurring problem of exceeding the upper-bound imposed by {\it Chandra} in the X-ray emission spectrum. We also note that the HARM simulation has a much smaller initially torus, which effectively reduces the bremsstrahlung X-ray flux. Although this itself may suggest that the angular momentum of the accretion flow must be low \citep{C99, R04}, one needs to explore the dependence of the bremsstrahlung X-ray emission on the size of the initial torus to quantify this constraint.

The above challenge of modeling observations of Sgr A* with GRMHD simulations may be alleviated by several means. First, as we have shown in the paper, nonthermal synchrotron emission is more efficient than thermal synchrotron and in the nonthermal scenario the radio spectrum may be fitted with a lower density normalization than the thermal one, which reduces the bremsstrahlung X-ray flux. The Doppler effect, which we didn't consider in the current study, is expected to boost the radio emission as well. Introducing a black hole spin, one can produce even lower values of plasma $\beta$ and higher Doppler boosting, both of which will enhance the radio emission. Lastly, the bremsstrahlung X-ray may be suppressed by reducing the mean density in the accretion disk through the tilt of the accretion disk with respect to the black hole spin \citep{Liu02b}. Three dimensional simulations are then needed to quantify this effect  \citep{R05, F07}.

In our modeling, we haven't studied the dependence of the model spectrum on the inclinational angle of the accretion disk, the GR effect, and the transfer of polarized emission. However, the radio spectrum is expected to have very weak dependence on these factors. Nevertheless, they should be included in a more complete study.
%The modeling of the second simulation appears promising and we believe that applying the nonthermal synchrotron emission model we can accurately match the observed spectra without exceeding the upper bound on the X-ray emission.  We also discovered that the magnetic fields produced in the simulations needed to be quite high in relation to the particle density of the accretion disc, meaning that the accretion disc $\beta$ factor must be quite low, less then one hundred. If we add spin to the current HARM code the magnetic field inside of the accretion disc will increase and it would be easier to match the observed spectra without exceeding the x-ray upper bound.  We also did not incorporate the Doppler effect, we suspect that this would slightly alter our final spectrum but again not drastically enough to fix the underlying problem.  We have also not yet included arbitrary line of sight or taken into account polarization, again this should not effect our results drastically by is needed for a more complete analysis of the model. Future, more complete models, would include all of these effects.

\section{Acknowledgments}

I would like to thank my mentor Siming Liu for guiding me throughout the duration of this project.  I would like to also thank Chris Fragile and Cong Yu for running the simulations.
%I would like to thank Chris Fryer for writing the simulations that we used to try and match the observed spectrum.  
I would like to also thank James Colgan and Norm Magee for all of their help during this summer.  Thanks to the NSF in part for funding.  I would finally like to thank UNM, Los Alamos and Sally Siedel for giving me the opportunity to participate in this great program.

\newpage
\end{document}